\documentclass[20pt]{article}
\usepackage{amsmath}
\usepackage{amssymb}
\usepackage{latexsym}
\usepackage{amsmath}
\usepackage{lineno}
\usepackage{graphicx}
\usepackage{float}
\usepackage{subfigure}
\usepackage{geometry}
\usepackage{graphics}
\usepackage{subfigure}
\usepackage{graphicx}
\usepackage{multirow}
\usepackage{booktabs}
\usepackage{hyperref}
\usepackage{cite}
\usepackage{natbib}
\makeatletter

\newcommand{\be}{\begin{equation}}
\newcommand{\ee}{\end{equation}}

\begin{document}
\begin{center}
\large {\bf Hawking Radiation of Vector Particles via Tunneling From 4-Dimensional And 5-Dimensional  Black Holes}
\end{center}

\begin{center}
Zhongwen Feng, $^1$
 $\footnote{E-mail:  zwfengphy@163.com}$
Yang Chen, $ ^{2}$
Xiaotao Zu ${^1}$
 $\footnote{E-mail:  xtzu@uestc.edu.cn}$
\end{center}

\begin{center}
\textit{1. School of Physical Electronics, University of Electronic Science and Technology of China, Chengdu, 610054, China\\
2. Institute of Theoretical Physics, China West Normal University, Nanchong, 637009, China}
\end{center}

\noindent
{\bf Abstract:} Using Proca equation and WKB approximation, we investigate Hawking radiation of vector particles via tunneling from 4-dimensional Kerr-de Sitter black hole and 5-dimensional Schwarzschild-Tangherlini black hole. The results show that the tunneling rates and Hawking temperatures are depended on the properties of spacetime (event horizon, mass and angular momentum). Besides, our results are the same as scalars and fermions tunneling from 4-dimensional Kerr-de Sitter black hole and 5-dimensional Schwarzschild-Tangherlini black hole.


\section{Introduction}
In 1970s, Bekenstein and Hawking discovered the thermodynamic of black holes, which indicates that black holes have thermal radiation \citep{ch1,ch2}. Later, Hawking investigated the radiation from black holes by quantum mechanics, and presented the theory of Hawking radiation. The existence of thermodynamic for black hole was an important discovery, which great influence on the foundations of physics. Thus, the Hawking radiation attracted researchers' attention.

The quantum tunneling for black holes is considered as an important method to study the Hawking radiation. This method was put forward by \cite{ch3}. Later, \cite{ch5,ch4} developed the quantum tunneling method and studied the Dirac particles tunneling from spherically symmetric black holes.  The Hamilton-Jacobi ansatz is another kind of tunneling method. Using the WKB approximation, the tunneling rate can be calculated by the formula $\Gamma  \propto \exp \left( { - 2{\mathop{\rm Im}\nolimits} S_0} \right)$, where $S_0$ is the classical action at the leading order in $\hbar$, then the Hawking temperature is obtained \citep{ch34,ch40}. Subsequently, higher order calculations of scalars, fermions and bosons tunneling from black holes are studied by \cite{ch6666+,ch6665+,ch5+,ch6+}. By defining a new of the particle energy, their work is consistent with the original results, which incidents there is no higher-order corrections to the Hawking temperature. Following that, a lot of work has been done for studying the Hawking radiation from black holes \citep{ch14+,ch15+,ch12,ch8,ch13,ch11,ch2+,ch56,ch18++,ch16+,ch7,ch10,ch17+,ch9,ch90}.

Obviously, the black holes do not only radiate scalar particles and Dirac particles, they may emit particles with arbitrary spin. Recently,  \cite{ch14++,ch14} and \cite{ch15} researched the vector (namely, spin-1 bosons) tunneling from black holes. However, all of their work are limited to the vector particles tunneling from the low dimensional (1+1 dimensional black hole and 1+2 dimensional) black holes. In this paper, the vector particles tunneling from 4-dimensional Kerr-de Sitter (KdS) black hole and 5-dimensional Schwarzschild-Tangherlini (ST) black hole are investigated with the help of the Proca equation and  WKB approximation. The KdS black hole is an important rotating solution of the Einstein equation with a positive cosmological constant. As we know, the cosmological constant is one kind of candidates for dark energy, it consistent with the current Lambda-CDM original model, and can explain the accelerating expansion of our universe. Besides, the KdS black holes have four horizons. Its special structure can help people further understand the properties of gravity. On the other hand, the 5-dimensional ST black hole is a good approximation to a 5 dimensional compactified black hole, people can use it to study the higher dimensional distorted compactified spacetime. From what has been discussed above, we think the vector tunneling from the 4-dimensional KdS black hole and the 5-dimensional ST black hole are worth to be studied, people may obtain more information of Hawking radiation from the higher-dimensional black holes.

The paper is organized as follows. In next section, extending the work of \cite{ch14++}, the vector tunneling from 4-dimensional KdS black hole will be investigated. In section 3, in the 5-dimensional ST spacetime, the spin-1 bosons tunneling radiation is derived. Section 4 is devoted to our discussions and conclusions.

\section{Vector tunneling from the 4-dimensional Kerr-de Sitter black hole}
Extending the method which presented by Kruglov, the vector tunneling from the 4-dimensional KdS black hole is investigated in this section. In \cite{ch16}, the metric of 4-dimensional KdS black hole is given by
\begin{eqnarray}
\label{eq1}
\begin{array}{l}
 ds^2  =  - \frac{\Delta }{{\rho ^2 }}\left( {dt - \frac{{a\sin ^2 \theta }}{\Xi }d\phi } \right)^2  + \rho ^2 \left( {\frac{1}{\Delta }dr^2  + \frac{1}{{\Delta _\theta  }}d\theta ^2 } \right) + \frac{{\Delta _\theta  \sin ^2 \theta }}{{\rho ^2 }}\left( {adt - \frac{{r^2  + a^2 }}{\Xi }d\phi } \right)^2 , \\
 \end{array}
\end{eqnarray}
where $\Delta  = \left( {r^2  + a^2 } \right)\left( {1 - r^2 l^{ - 2} } \right) - 2Mr$, $\Delta _\theta   = 1 + a^2 l^{ - 2} \cos ^2 \theta$, $\Xi  = 1 + a^2 l^{ - 2}$ and $\rho ^2  = r^2  + a^2 \cos ^2 \theta$, $M$ and $a$ are the mass and angular momentum of 4-dimensional KdS black hole, $l$ is a constant depended on the cosmological factor as $\Lambda=3 l^{-2}$,  respectively. One can obtains four horizons when $\Delta=0$. The four horizons are outer cosmological horizon $r_{c+}$, outer event horizon $r_{+}$, inner event horizon $r_{-}$ and inner cosmological horizon $r_{c-}$, which satisfy the relation $r_{c+}>r_{+}>r_{-}>0>r_{c-}$. For convenience, we redefined Eq.(\ref{eq1}) as
\begin{eqnarray}
\label{eq3}
\begin{array}{l}
 ds^2  =  - A( r )dt^2  + B( r )^{ - 1} dr^2  + C( r )d\theta ^2 + D( r )d\phi ^2 + 2E( r )dtd\phi,  \\
 \end{array}
\end{eqnarray}
the notations are $A\left( r \right) = \Delta  - a\Delta _\theta  \sin ^2 \theta /\rho ^2$, $B\left( r \right) = \Delta /\rho ^2$, $C\left( r \right) = \rho ^2 /\Delta _\theta$, $D\left( r \right) = \sin ^2 \theta [\Delta _\theta  (a^2  + r^2 )^2  - a^2 \Delta \sin ^2 \theta ]/\Xi ^2 \rho ^2$, and $
E\left( r \right) = a\sin ^2 \theta [\Delta  + (a^2  + r^2 )]\Delta _\theta  /\Xi \rho$, respectively.

According to \cite{ch33,ch14++,ch14}, the dynamic behavior of vector particles in curved spacetime is described by Proca equation
\begin{eqnarray}
\label{eq6}
D_\mu  \psi ^{ \nu \mu }  + \frac{{m^2 }}{{\hbar ^2 }}\psi ^\nu   = 0,
\end{eqnarray}
\begin{eqnarray}
\label{eq7}
\psi _{\nu \mu}  = D_\nu  \psi _\mu   - D_\mu  \psi _\nu   = \partial _\nu  \psi _\mu   - \partial _\mu  \psi _\nu  ,
\end{eqnarray}
where $D_\mu$ are covariant derivatives, $\psi _\nu$ is related to the $\psi _t$, $\psi _r$, $\psi _\theta$ and  $\psi _\phi$, $m$ is the mass of vector particles. $\psi ^{\mu \nu }$  is an anti-symmetrical tensor. Thus, the Eq. (\ref{eq6}) can be rewritten as follows
\begin{eqnarray}
\label{eq8}
\frac{1}{{\sqrt { - g} }}\partial _\mu  \left( {\sqrt { - g} \psi ^{\nu \mu } } \right) + \frac{{m^2 }}{{\hbar ^2 }}\psi ^\nu   = 0.
\end{eqnarray}
The components  of $\psi ^\nu$  and $\psi ^{\mu \nu }$  are
\begin{eqnarray}
\label{eq9}
\begin{array}{l}
\begin{array}{*{20}c}
   {\psi ^0  = \frac{{ - D\psi _0  + E\psi _3 }}{{AD + E^2 }},} & {\psi ^1  = B\psi _1 ,} & {\psi ^2  = C^{ - 1} \psi _2 ,}  \\
\end{array}
\\
 \\
\begin{array}{*{20}c}
   {\psi ^3  = \frac{{E\psi _0  + A\psi _3 }}{{AD + E^2 }},} & {\psi ^{01}  = \frac{{ - DB\psi _{01}  - EB\psi _{13} }}{{AD + E^2 }},}  \\
\end{array}
\\
 \\
\begin{array}{*{20}c}
{\psi ^{02}  = \frac{{C^{ - 1} \left( { - D\psi _{02}  - E\psi _{23} } \right)}}{{AD + E^2 }},} & {\psi ^{03}  =  - \frac{{\psi _{03} }}{{AD + E^2 }},}  \\
\end{array}
\\
 \\
\begin{array}{*{20}c}
   {\psi ^{12}  = BC^{ - 1} \psi _{12} ,} & {\psi ^{13}  = \frac{{B\left( {A\psi _{13}  - E\psi _{01} } \right)}}{{AD + E^2 }},}  \\
\end{array}
\\
\\
  \psi ^{23}  = \frac{{C^{ - 1} \left( {A\psi _{23}  - E\psi _{02} } \right)}}{{AD + E^2 }}.
 \end{array}
\end{eqnarray}
Substituting Eq. (\ref{eq7}) and Eq. (\ref{eq9}) into Eq. (\ref{eq6}), one yields
\begin{eqnarray}
\label{eq10}
\begin{array}{l}
 \frac{1}{{\sqrt { - g} }}\left\{ {\partial _r \left[ {\sqrt { - g} \left( {\frac{{ - DB\psi _{01}  - EB\psi _{13} }}{{AD + E^2 }}} \right)} \right]} \right. +
 \partial _\theta  \left[ {\sqrt { - g} \left( {\frac{{ - DC^{ - 1} \psi _{02}  - EC^{ - 1} \psi _{23} }}{{AD + E^2 }}} \right)} \right] + \\
 \\
 \left. {\partial _\phi  \left[ {\sqrt { - g} \left( { - \frac{{\psi _{03} }}{{AD + E^2 }}} \right)} \right]} \right\} + \frac{{m^2 }}{{\hbar ^2 }}\left( {\frac{{E\psi _3  - D\psi _0 }}{{AD + E^2 }}} \right) = 0, \\
 \end{array}
\end{eqnarray}
\begin{eqnarray}
\label{eq11}
\begin{array}{l}
 \frac{1}{{\sqrt { - g} }}\left\{ {\partial _t \left[ {\sqrt { - g} \left( {\frac{{DB\psi _{01}  + EB\psi _{13} }}{{AD + E^2 }}} \right)} \right]} \right.
  + \partial _\theta  \left( {\sqrt { - g} BC^{ - 1} \psi _{12} } \right) \\
  \\
  + \left. {\partial _\phi  \left[ {\sqrt { - g} \left( {\frac{{AB\psi _{13}  - BE\psi _{01} }}{{AD + E^2 }}} \right)} \right]} \right\} + \frac{{m^2 }}{{\hbar ^2 }}B\psi _1  = 0, \\
 \end{array}
\end{eqnarray}
\begin{eqnarray}
\label{eq12}
\begin{array}{l}
 \frac{1}{{\sqrt { - g} }}\left\{ {\partial _t \left[ {\sqrt { - g} \left( {\frac{{DC^{ - 1} \psi _{02}  + EC^{ - 1} \psi _{23} }}{{AD + E^2 }}} \right)} \right]} \right.
  + \partial _\theta  \left[ {\sqrt { - g} \left( { - BC^{ - 1} \psi _{12} } \right)} \right] \\
  \\
 \left. { + \partial _\phi  \left[ {\sqrt { - g} \left( {\frac{{C^{ - 1} A\psi _{23}  - C^{ - 1} E\psi _{02} }}{{AD + E^2 }}} \right)} \right]} \right\}  + \frac{{m^2 }}{{\hbar ^2 }}C^{ - 1} \psi _2  = 0, \\
 \end{array}
\end{eqnarray}
\begin{eqnarray}
\label{eq13}
\begin{array}{l}
 \frac{1}{{\sqrt { - g} }}\left\{ {\partial _t \left[ {\sqrt { - g} \left( {\frac{{\psi _{03} }}{{AD + E^2 }}} \right)} \right]} \right.
  + \partial _r \left[ {\sqrt { - g} \left( {\frac{{BE\psi _{01}  - AB\psi _{13} }}{{AD + E^2 }}} \right)} \right] \\
  \\
  + \left. {\partial _\theta  \left[ {\sqrt { - g} \left( {\frac{{C^{ - 1} E\psi _{02}  - C^{ - 1} A\psi _{23} }}{{AD + E^2 }}} \right)} \right]} \right\}
  + \frac{{m^2 }}{{\hbar ^2 }}\left( {\frac{{E\psi _0  + A\psi _3 }}{{AD + E^2 }}} \right) = 0. \\
 \end{array}
\end{eqnarray}
For solving Eqs. (\ref{eq10}) - (\ref{eq13}),  $\psi _\nu$ taking form as

\begin{eqnarray}
\begin{aligned}
\label{eq14}
{\psi _\nu   = \left( {c_\nu  } \right)\exp \left[ {\frac{i}{\hbar }S_0 \left( {t,r,\theta ,\phi } \right) + \sum\limits_n {\hbar ^n } S_n \left( {t,r,\theta ,\phi } \right)} \right]},
\end{aligned}
\end{eqnarray}
where $n=1,2,3,\cdots$. Since the WKB approximation is been applied here, the higher order terms of $\mathcal{O}(\hbar)$ are neglected. The resulting equations to leading order in $\hbar$ are
\begin{eqnarray}
\label{eq15}
\begin{array}{l}
 DB\left[ {c_1 \left( {\partial _t S} \right)\left( {\partial _r S} \right) - c_0 \left( {\partial _r S} \right)^2 } \right]
  + EB\left[ {c_3 \left( {\partial _r S} \right)^2  - c_1 \left( {\partial _\phi  S} \right)\left( {\partial _r S} \right)} \right] \\
  \\
  + DC^{ - 1} \left[ {c_2 \left( {\partial _t S} \right)\left( {\partial _\theta  S} \right) - c_0 \left( {\partial _\theta  S} \right)^2 } \right]
  + EC^{ - 1} \left[ {c_3 \left( {\partial _\theta  S} \right)^2  - c_2 \left( {\partial _\phi  S} \right)\left( {\partial _\theta  S} \right)} \right] \\
  \\
  + \left[ {c_3 \left( {\partial _\phi  S} \right)\left( {\partial _t S} \right) - c_0 \left( {\partial _\phi  S} \right)^2 } \right] + m^2 \left( {c_3 E - c_0 D} \right) = 0, \\
 \end{array}
\end{eqnarray}
\begin{eqnarray}
\label{eq16}
\begin{array}{l}
 DB\left[ {c_0 \left( {\partial _r S} \right)\left( {\partial _t S} \right) - c_1 \left( {\partial _t S} \right)^2 } \right]
  + EB\left[ {c_3 \left( {\partial _t S} \right)\left( {\partial _r S} \right)c_1 \left( {\partial _\phi  S} \right)\left( {\partial _t S} \right)} \right] \\
  \\
  + BC^{ - 1} \left( {AD + E^2 } \right)\left[ {c_1 \left( {\partial _\theta  S} \right)^2  - c_2 \left( {\partial _\theta  S} \right)\left( {\partial _r S} \right)} \right]  + BE\left[ {c_1 \left( {\partial _t S} \right)\left( {\partial _\varphi  S} \right) - c_0 \left( {\partial _r S} \right)\left( {\partial _\phi  S} \right)} \right] \\
  \\
  + AB\left[ {c_1 \left( {\partial _\phi  S} \right)^2  - c_3 \left( {\partial _r S} \right)\left( {\partial _\phi  S} \right)} \right] + c_1 m^2 B\left( {AD + E^2 } \right) = 0, \\
 \end{array}
\end{eqnarray}
\begin{eqnarray}
\label{eq17}
\begin{array}{l}
 DC^{ - 1} \left[ {c_0 \left( {\partial _t S} \right)\left( {\partial _\theta  S} \right) - c_2 \left( {\partial _t S} \right)^2 } \right]
  + EC^{ - 1} \left[ {c_3 \left( {\partial _t S} \right)\left( {\partial _\theta  S} \right) - c_2 \left( {\partial _t S} \right)\left( {\partial _\phi  S} \right)} \right] \\
  \\
  + BC^{ - 1} \left( {AD + E^2 } \right)\left[ {c_2 \left( {\partial _r S} \right)^2  - c_1 \left( {\partial _r S} \right)\left( {\partial _\theta  S} \right)} \right]  + C^{ - 1} E\left[ {c_2 \left( {\partial _t S} \right)\left( {\partial _\phi  S} \right) - c_0 \left( {\partial _\phi  S} \right)\left( {\partial _\theta  S} \right)} \right] \\
  \\
  + C^{ - 1} A\left[ {c_2 \left( {\partial _\phi  S} \right)^2  - c_3 \left( {\partial _\theta  S} \right)\left( {\partial _\phi  S} \right)} \right] + c_2 m^2 C^{ - 1} \left( {AD + E^2 } \right) = 0, \\
 \end{array}
\end{eqnarray}
\begin{eqnarray}
\label{eq18}
\begin{array}{l}
 \left[ {c_0 \left( {\partial _\phi  S} \right)\left( {\partial _t S} \right) - c_3 \left( {\partial _t S} \right)^2 } \right] + BE\left[ {c_0 \left( {\partial _r S} \right)^2  - c_1 \left( {\partial _t S} \right)\left( {\partial _r S} \right)} \right] \\
  \\
  + AB\left[ {c_1 \left( {\partial _r S} \right)\left( {\partial _\phi  S} \right) - c_3 \left( {\partial _r S} \right)^2 } \right]  + C^{ - 1} E\left[ {c_0 \left( {\partial _\theta  S} \right)^2  - c_2 \left( {\partial _\theta  S} \right)\left( {\partial _t S} \right)} \right] \\
  \\
  + C^{ - 1} A\left[ {c_2 \left( {\partial _\theta  S} \right)\left( {\partial _\phi  S} \right) - c_3 \left( {\partial _\theta  S} \right)^2 } \right] + m^2 \left( {c_0 E + c_3 A} \right) = 0. \\
 \end{array}
\end{eqnarray}
Considering the spacetime of metric (\ref{eq1}) has two Killing vectors ${\partial _ t}$ and ${\partial _\phi}$, the solutions of Eqs. (\ref{eq15}) - (\ref{eq18}) are in the form
\begin{eqnarray}
\label{eq19}
S_0  =  -\omega t + W\left( r \right)+ \Theta \left( \theta  \right) + j\phi  ,
\end{eqnarray}
where  $\omega$ and  $j$  are the energy and the angular momentum of vector particles, respectively. Putting Eq. (\ref{eq19}) into Eqs. (\ref{eq15}) - (\ref{eq18}), one gets
\begin{eqnarray}
\label{eq20}
\Lambda \left( {c_0 ,c_1 ,c_2 ,c_3 } \right)^T  = 0.
\end{eqnarray}
The  $\Lambda $ in Eq. (\ref{eq20}) is a $4\times4$ matrix, its components are
\begin{equation}
\begin{aligned}
\label{eq21}
\begin{array}{l}
 \Lambda _{00} \! = \!{ - Dm^2  - j^2  - BD\left( {W'} \right)^2  - DC^{ - 1} \left( {\partial _\theta  \Theta } \right)^2 }, \Lambda _{01} \! =\! { - BEW'j - BDW'\omega },\\
 \\
  \Lambda _{02}  \!= \!{ - C^{ - 1} Ej\left( {\partial _\theta  \Theta } \right) - C^{ - 1} D\left( {\partial _\theta  \Theta } \right)\omega }, \Lambda _{03} \! = \!{BE\left( {W'} \right)^2  + Em^2  + C^{ - 1} E\left( {\partial _\theta  \Theta } \right)^2  - j\omega },\\
 \\
 \Lambda _{10} \! = \! { - BEW'\left( {\partial _\theta  \Theta } \right) - BDW'\omega },  \Lambda _{11} \! = \! m^2 B\left( {AD + E^2 } \right) - DB\omega ^2  - EBj\omega  - C^{ - 1} B\left( {\partial _\theta  \Theta } \right)^2 - EB\left( {\partial _\theta  \Theta } \right)\omega  - ABj^2 ,\\
 \\
\Lambda _{12} \! =  \! - BC^{ - 1} W'\left( {\partial _\theta  \Theta } \right)\left( {AD + E^2 } \right), \Lambda _{13} \! = \!{BEW'\omega  - ABW'j}, \Lambda _{20}\!  = \! {EC^{ - 1} j\left( {\partial _\theta  \Theta } \right) - C^{ - 1} D\left( {\partial _\theta  \Theta } \right)\omega },
 \\
 \\
 \Lambda _{21}\! = \! - BC^{ - 1} \left( {\partial _\theta  \Theta } \right)W'\left( {AD + E^2 } \right),  \Lambda _{22} \! =\! m^2 C^{ - 1} \left( {AD + E^2 } \right) - C^{ - 1} D\omega ^2  - C^{ - 1} Ej\omega\\
  \\
 \!   -   \!  BC^{ - 1} \left( {W'} \right)^2  - AC^{ - 1} j^2  - C^{ - 1} EjW', \Lambda _{23}\!  = \!{C^{ - 1} E\left( {\partial _\theta  \Theta } \right)\omega  - A\left( r \right)C\left( r \right)^{ - 1} j\left( {\partial _\theta  \Theta } \right)},
 \\
 \\
  \Lambda _{30} \! =\! {m^2 E - j\omega  + BE\left( {W'} \right)^2  + EC^{ - 1} \left( {\partial _\theta  \Theta } \right)^2 },  \Lambda _{31} \! =\! {BEW'\omega  - ABjW'},\\
  \\
   \Lambda _{32}\! = \!{EC^{ - 1} \left( {\partial _\theta  \Theta } \right)\omega  - AC^{ - 1} j\Theta },  \Lambda _{33} \! = \!{Am^2  - \omega ^2  + AB\left( {W'} \right)^2  + AC^{ - 1} \left( {\partial _\theta  \Theta } \right)^2 },
 \end{array}
 \end{aligned}
\end{equation}
where $W' = \partial _r S_0$ and $j={\partial _\phi  S_0 } $. Eq. (\ref{eq20}) has a nontrivial solution when ${\rm{Det}}\left( \Lambda  \right) = 0$. Hence, one yields
\begin{eqnarray}
\label{eq22}
W_ \pm   = \int {\sqrt {\frac{{\left( {\omega ^2  - \Omega j} \right)^2  + X}}{{\left( {AD + E^2 } \right)BD^{ - 1} }}} } dr =   \pm i\pi\frac{{ {\omega  - \Omega \left( {r_ +  } \right)j}  }}{{2\kappa \left( {r_ +  } \right)}},
\end{eqnarray}
where $+ (-)$  are the outgoing (incoming) solutions on the outer event horizon, $X=-D^{ - 2} [E^2 j^2  + m^2 D({AD + E^2 } )]+ C^{ - 1} D^{ - 1} [( {AD + E^2 } ) - AC]$, $\Omega (r_ +  ) =-E(r_ +)/D(r_ +)=a\Xi /\left( {r_ + ^2  + a^2 } \right)$ is the angular velocity on outer the event
horizon and $\kappa ( {r_ +  }) =r_ +  (1 - 2r_ + ^2 l^{ - 2}  - r_ +  a^2 l^{ - 2} )/(r_ + ^2  + a^2)$ is the surface gravity of outer event horizon. The tunneling rate of vector particle from 4-dimensional KdS black hole is obtained as
\begin{eqnarray}
\label{eq23}
\Gamma  =\frac{{\Gamma _{\left( {emission} \right)} }}{{\Gamma _{\left( {absorpation} \right)} }} = \frac{{\exp \left( { - 2{\mathop{\rm Im}\nolimits} W_ +   - 2{\mathop{\rm Im}\nolimits} \Xi } \right)}}{{\exp \left( { - 2{\mathop{\rm Im}\nolimits} W_ -   - 2{\mathop{\rm Im}\nolimits} \Xi } \right)}}.
\end{eqnarray}
As we known, any particle outside the event horizon will fall into the black hole, one has  $\Gamma _{absorpation}  = 1$, namely, ${\rm{ImW}}_ -   + {\mathop{\rm Im}\nolimits} \Xi  = 0$. Therefore, the result is
\begin{eqnarray}
\label{eq24}
\Gamma  = \exp \left[ { - 2\pi \frac{{\omega  - j\Omega \left( {r_ +  } \right)}}{{\kappa \left( {r_ +  } \right)}}} \right].
\end{eqnarray}
 With the help of Boltzman factor \cite{ch18}, the Hawking temperature of 4-dimensional KdS black hole is
\begin{eqnarray}
\label{eq25}
T_{ H }  = \frac{{r_{ + }  - r_{ + }^3 l^{ - 2}  - r_{ + } a^2 l^{ - 2}  - M}}{{2\pi \left( {r_{ + }^2  + a^2 } \right)}}.
\end{eqnarray}
 Eq. (\ref{eq24}) and Eq. (\ref{eq25}) are the vector particles tunneling rate and Hawking temperature of KdS black hole. The results showed that the $\Gamma$  and  $T_H $ are depended on the outer event horizon $r_{+}$, mass $M$, angular momentum of KdS black hole $a$ and the cosmological factor $\Lambda$. When $\Lambda=0$, the Eq. (\ref{eq25}) is reduced to the temperature of Kerr black hole. When $a=0$, $\Lambda=0$ and $r_+=2M$, the Hawking temperature of Schwarzschild (SC) black hole $T_{H\left( {sc} \right)}  = {1 \mathord{\left/ {\vphantom {1 {8\pi M}}} \right. \kern-\nulldelimiterspace} {8\pi M}}$ is recovered \citep{ch0,ch0+}. Moreover, our result is fully in accordance with that obtained by other methods \citep{ch31,ch30,ch11,ch17+,ch32}.

\section{Vector tunneling from the 5-dimensional Schwarzschild-Tangherlini black hole}
In this section, we study the quantum tunneling from 5 dimensional Schwarzschild-Tangherlini black hole via Proca equation. When added extra compact spatial dimensions into a static spherically symmetric solution of the vacuum Einstein equations, one gets the line element of $D$-dimensional ST black hole \cite{ch18+}
\begin{eqnarray}
\label{eq26}
ds^2  =  - f\left( r \right)dt^2  + f\left( r \right)^{ - 1} dr^2  + r^2 d\Omega _{D-2}^2,
\end{eqnarray}
where $f\left( r \right) = 1 - \left( {{{r_H } \mathord{\left/ {\vphantom {{R_H } r}} \right. \kern-\nulldelimiterspace} r}} \right)^{D - 3}$, $r_H$ marks the event horizon of the $D$-dimensional ST black hole, which is related to the mass of ST black hole as $ M = {{\left( {D - 3} \right)r_H^{D - 3} } \mathord{\left/
 {\vphantom {{\left( {D - 3} \right)r_H^{D - 3} } 2}} \right.\kern-\nulldelimiterspace} 2}$, $d\Omega _{D - 2}^2  = d\chi _2^2  + \sin ^2 \chi _2 d\chi _3^2  +  \cdots +\left( {\prod\limits_{k = 2}^{D - 2} {\sin ^2 \chi _k } } \right)d\chi _{D - 1}^2$ is the line element on $D-2$-sphere with the  angles on this sphere ${\chi _k }$.  For $D = 5$, one has \cite{ch19+}
\begin{eqnarray}
\label{eq26+}
\begin{array}{l}
 ds^2  =  - \widetilde{A}\left( r \right)dt^2  + \widetilde{B}^{ - 1} \left( r \right)dr^2  + \widetilde{C}\left( r \right)d\zeta ^2 + \widetilde{D}\left( r \right)d\vartheta ^2  + \widetilde{E}\left( r \right)d\phi ^2
  \\
 \\
  =
 - \left( {1 - {{r_H^2 } \mathord{\left/ {\vphantom {{r_H^2 } {r^2 }}} \right. \kern-\nulldelimiterspace} {r^2 }}} \right)dt^2  + \left( {1 - {{r_H^2 } \mathord{\left/ {\vphantom {{r_H^2 } {r^2 }}} \right. \kern-\nulldelimiterspace} {r^2 }}} \right)^{ - 1} dr^2 + r^2 d\zeta ^2  + r^2 \sin ^2 \zeta d\vartheta ^2 + r^2 \sin ^2 \zeta \sin ^2 \vartheta d\phi ^2,  \\
 \end{array}
\end{eqnarray}
where $r_H^2  \sim M$ \cite{ch19++}. Near the event horizon, it is clear that $\widetilde{B}\left( r \right) = \widetilde{B}'\left( {r_H } \right)\left( {r - r_H } \right) +  \mathcal{O}\left[ {\left( {r - r_H } \right)^2 } \right]$. According to Eq. (\ref{eq7}), one yields
\begin{eqnarray}
\label{eq27}
\begin{array}{l}
\begin{array}{*{20}c}
   {\psi ^0 \! =  \!- \widetilde{A}^{ - 1} \psi _0 ,} & {\psi ^1 \! =\! \widetilde{B}\psi _1 ,} & {\psi ^2 \! = \!\widetilde{C}^{ - 1} \psi _2 ,}  \\
\end{array}
 \\
 \\
\begin{array}{*{20}c}
   {\psi ^3 \! = \!\widetilde{D}^{ - 1} \psi _3 ,} & {\psi ^4  = \widetilde{E}^{ - 1} \psi _4 ,} & {\psi ^{01}\!  = \! - \widetilde{B}\widetilde{A}^{ - 1} \psi _{01} ,}  \\
\end{array}
\\
 \\
\begin{array}{*{20}c}
   {\psi ^{02}\!  = \! - (\widetilde{A}\widetilde{C})^{ - 1} \psi _{02} ,} & {\psi ^{03} \! =  \!- (\widetilde{A}\widetilde{D})^{ - 1} \psi _{03} ,}  \\
\end{array}
 \\
 \\
\begin{array}{*{20}c}
   {\psi ^{04} \! = \! - (\widetilde{A}\widetilde{E})^{ - 1} \psi _{04} ,} & {\psi ^{12} \! = \!{\rm{ }}\widetilde{B}\widetilde{C}^{ - 1} \psi _{12} ,}  \\
\end{array}
 \\
\\
\begin{array}{*{20}c}
   {\psi ^{13}\!  = \!\widetilde{B}\widetilde{D}^{ - 1} \psi _{13} ,} & {\psi ^{14}\!  = \!\widetilde{B}\widetilde{E}^{ - 1} \psi _{14} ,}  \\
\end{array}
\\
\\
\begin{array}{*{20}c}
   {\psi ^{23} \! = \!(\widetilde{C}\widetilde{D})^{ - 1} \psi _{23} ,} & {\psi ^{24}\!  = \!(\widetilde{C}\widetilde{E})^{ - 1} \psi _{24} ,}  \\
\end{array}
\\
 \\
 \psi ^{34} \! = \!{{(\widetilde{D}\widetilde{E})}^{-1}}\psi _{34}.\\
 \end{array}
\end{eqnarray}
Inserting Eq. (\ref{eq27}) and $\psi _\nu   = \left( {c_\nu } \right)\exp[ {\frac{i}{\hbar }S_0 \left( {t,r,\zeta ,\vartheta ,\phi } \right)} ]$ into Eq. (\ref{eq6}), and ignoring the higher order of terms of $\hbar$, we have
\begin{eqnarray}
\label{eq28}
\begin{array}{l}
\widetilde{{B}}\left[ {c_0 \left( {\partial _r S_0 } \right)^2  - c_1 \left( {\partial _r S_0 } \right)\left( {\partial _t S_0 } \right)} \right]+ \widetilde{{C}}^{ - 1} \left[ {c_0 \left( {\partial _\zeta  S_0 } \right)^2  - c_2 \left( {\partial _\zeta  S_0 } \right)\left( {\partial _t S_0 } \right)} \right]+ \\
 \\
 \widetilde{{D}}^{ - 1} \left[ {c_0 \left( {\partial _\vartheta  S_0 } \right)^2  - c_3 \left( {\partial _\vartheta  S_0 } \right)\left( {\partial _t S_0 } \right)} \right]+ \widetilde{{E}}^{ - 1} \left[ {c_0 \left( {\partial _\phi  S_0 } \right)^2  - c_4 \left( {\partial _\phi  S_0 } \right)\left( {\partial _t S_0 } \right)} \right] \\
 \\
  + c _0 m^2   = 0, \\
 \end{array}
 \end{eqnarray}
\begin{eqnarray}
\label{eq29}
\begin{array}{l}
{\widetilde{{A}} ^{-1}}\left[ {c_0 \left( {\partial _r S_0} \right)\left( {\partial _t S_0} \right) - c_1 \left( {\partial _t S_0} \right)^2 } \right]+ {\widetilde{{C}} ^{-1}}\left[ {c_1 \left( {\partial _\zeta  S_0} \right)^2  - c_2 \left( {\partial _r S_0} \right)\left( {\partial _\zeta  S_0} \right)} \right] \\
 \\
  + {\widetilde{{D}} ^{-1}}\left[ {c_1 \left( {\partial _\vartheta  S_0} \right)^2  - c_3 \left( {\partial _r S_0} \right)\left( {\partial _\vartheta  S_0} \right)} \right]+ {\widetilde{{E}} ^{-1}}\left[ {c_1 \left( {\partial _\phi S_0} \right)^2  - c_4 \left( {\partial _r S_0} \right)\left( {\partial _\phi S_0} \right)} \right] \\
  \\
  +c _1 m^2  = 0, \\
 \end{array}
 \end{eqnarray}
\begin{eqnarray}
\label{eq30}
\begin{array}{l}
 {\widetilde{{A}} ^{-1}}\left[ {c_0 \left( {\partial _\zeta S_0} \right)\left( {\partial _t S_0} \right) - c_2 \left( {\partial _t S_0} \right)^2 } \right] + \widetilde{{B}}\left[ {c_2 \left( {\partial _r S_0} \right)^2  - c_1 \left( {\partial _\zeta  S_0} \right)\left( {\partial _r S_0} \right)} \right] \\
 \\
  + {\widetilde{{D}} ^{-1}}\left[ {c_2 \left( {\partial _\vartheta  S_0} \right)^2  - c_3 \left( {\partial _\zeta  S_0} \right)\left( {\partial _\vartheta  S_0} \right)} \right]+ {\widetilde{{E}} ^{-1}}\left[ {c_2 \left( {\partial _\phi S_0} \right)^2  - c_4 \left( {\partial _\zeta  S_0} \right)\left( {\partial _\phi S_0} \right)} \right] \\
  \\
  +c _2 m^2   = 0, \\
 \end{array}
 \end{eqnarray}
\begin{eqnarray}
\label{eq31}
\begin{array}{l}
 {\widetilde{{A}} ^{-1}}\left[ {c_0 \left( {\partial _\vartheta  S_0} \right)\left( {\partial _t S_0} \right) - c_3 \left( {\partial _t S_0} \right)^2 } \right] + \widetilde{{B}}\left[ {c_3 \left( {\partial _r S_0} \right)^2  - c_1 \left( {\partial _r S_0} \right)\left( {\partial _\vartheta  S_0} \right)} \right] \\
 \\
  + {\widetilde{{C}} ^{-1}}\left[ c_3 \left( {\partial _\vartheta  S_0} \right)^2 - { c_2 \left( {\partial _\vartheta  S_0} \right)\left( {\partial _\zeta  S_0} \right) } \right] +{\widetilde{{E}} ^{-1}}\left[ {c_3 \left( {\partial _\phi S_0} \right)^2  - c_4 \left( {\partial _\phi S} \right)\left( {\partial _\vartheta  S} \right)} \right] \\
  \\
  +c _3 m^2   = 0, \\
 \end{array}
 \end{eqnarray}
\begin{eqnarray}
\label{eq32}
\begin{array}{l}
 {\widetilde{{A}} ^{-1}}\left[ {c_0 \left( {\partial _\phi S_0} \right)\left( {\partial _t S_0} \right) - c_4 \left( {\partial _t S_0} \right)^2 } \right]+ \widetilde{{B}}\left[ {c_4 \left( {\partial _r S_0} \right)^2  - c_1 \left( {\partial _r S_0} \right)\left( {\partial _\phi S_0} \right)} \right] \\
 \\
  + {\widetilde{{C}} ^{-1}}\left[ {c_4 \left( {\partial _\zeta  S_0} \right)^2  - c_2 \left( {\partial _\zeta  S_0} \right)\left( {\partial _\phi S_0} \right)} \right]+ {\widetilde{{D}} ^{-1}}\left[ {c_4 \left( {\partial _\vartheta  S_0} \right)^2  - c_3 \left( {\partial _\vartheta  S_0} \right)\left( {\partial _\phi S_0} \right)} \right] \\
  \\
  + c _4 m^2  = 0. \\
 \end{array}
 \end{eqnarray}
Considering property of spacetime, we carry out the separation of variables as
\begin{eqnarray}
\label{eq33+}
S_0 =  - \omega t + W\left( r \right) + \Theta \left( {\zeta ,\vartheta } \right) + j\phi,
\end{eqnarray}
where $\omega$  and  $j$ are the energy and angular momentum of the emitting particles. Putting Eq. (\ref{eq33+}) into Eqs. (\ref{eq28}) - (\ref{eq32}), one obtains a matrix equation  $\Lambda \left( {c_0 ,c_1 ,c_2 ,c_3 ,c_4 } \right)^T $. The components of $5\times5$ matrix $\Lambda$  are
\begin{eqnarray}
\label{eq33}
\begin{array}{l}
 \Lambda _{00}  = m^2 + \widetilde{{B}}\left( {W'} \right)^2  +{\widetilde{{C}} ^{-1}}\left( {\partial _\zeta  \Theta } \right)^2  +{\widetilde{{D}} ^{-1}}\left( {\partial _\vartheta  \Theta } \right)^2  \\
 \\\begin{array}{*{20}c}
{ + \widetilde{E}^{ - 1} j^2 ,} & {\Lambda _{01} \! =\! \widetilde{B} W'\omega ,} & {\Lambda _{02} \! = \!\widetilde{C}^{ - 1} \omega \left( {\partial _\zeta  \Theta } \right),}  \\
\end{array}
\\
 \\
\begin{array}{*{20}c}
{\Lambda _{03} \! = \! D^{ - 1} \omega \left( {\partial _\vartheta  \Theta } \right),} & {\Lambda _{04} \! =  \!E^{ - 1} \omega j,}  \\
\end{array}
\\
 \\
\Lambda _{10} \! = \! - {\widetilde{{A}} ^{-1}}\omega W',  \\
\\
\Lambda _{11} \! = \! m^2  - {\widetilde{{A}} ^{-1}}\omega ^2  + {\widetilde{{C}} ^{-1}}\left( {\partial _\zeta  \Theta } \right)^2   \\
 \\+ {\widetilde{{D}} ^{-1}}\left( {\partial _\vartheta \Theta } \right)^2 + {\widetilde{{E}} ^{-1}}j^2 ,\Lambda _{12}\!  = \! -{\widetilde{{C}} ^{-1}}W' \left( {\partial _\zeta  \Theta } \right), \\
 \\
 \Lambda _{13}\!  =\! -{\widetilde{{D}} ^{-1}}\left( {\partial _\vartheta  \Theta } \right)W' , \Lambda _{14}\!  = \! -{\widetilde{{E}} ^{-1}} j W',\\
 \\
\Lambda _{20} \! = \! - {\widetilde{{A}} ^{-1}}\omega \left( {\partial _\zeta  \Theta } \right), \Lambda _{21}  = -\widetilde{{B}}W'\left( {\partial _\zeta  \Theta } \right),\\
 \\
 \Lambda _{22} \! = \!m^2  - {\widetilde{{A}} ^{-1}}\omega ^2  + \widetilde{{B}}\left( {W'} \right)^2 + {\widetilde{{D}} ^{-1}}\left( {\partial _\vartheta \Theta } \right)^2 \\
  \\ + {\widetilde{{E}} ^{-1}}j^2,  \Lambda _{23}\!  =  \!- {\widetilde{{D}} ^{-1}}\left( {\partial _\vartheta \Theta } \right)\left( {\partial _\zeta  \Theta } \right),\\
 \\
 \Lambda _{24} \! = \! -{\widetilde{{E}} ^{-1}}j \left( {\partial _\zeta  \Theta } \right),\Lambda _{30} \! = \! - {\widetilde{{A}} ^{-1}}\omega \left( {\partial _\vartheta  \Theta } \right),\\
 \\
 \Lambda _{31} \! = \! -\widetilde{{B}} W'\left( {\partial _\vartheta  \Theta } \right), \Lambda _{32} \! = \! - {\widetilde{{C}} ^{-1}}\left( {\partial _\vartheta  \Theta } \right)\left( {\partial _\zeta  \Theta } \right),  \\
 \\
 \Lambda _{33}\!  = \!m^2  - {\widetilde{{A}} ^{-1}}\omega ^2  + \widetilde{{B}}\left( {W'} \right)^2  + {\widetilde{{C}} ^{-1}}\left( {\partial _\vartheta  \Theta } \right)^2 \\
 \\
\begin{array}{*{20}c}
   { + E^{ - 1} j^2 ,} & {\Lambda _{34}  =  - E^{ - 1} \left( {\partial _\vartheta  \Theta } \right)j,}  \\
\end{array}
\\
 \\
\begin{array}{*{20}c}
{\Lambda _{40}  =  - \widetilde{A}^{ - 1} \omega j,} & {\Lambda _{41}  =  - \widetilde{B} W'j,}  \\
\end{array}
\\
\\
\Lambda _{42}\!  =\!  - {\widetilde{{C}} ^{-1}}\left( {\partial _\zeta \Theta } \right) j, \Lambda _{43}  =  - {\widetilde{{D}} ^{-1}}j\left( {\partial _\vartheta \Theta } \right), \\
\\
\Lambda _{44} \!= \! m^2  - {\widetilde{{A}} ^{-1}}\omega ^2  + \widetilde{{B}}\left( {W'} \right)^2  + {\widetilde{{C}} ^{-1}}\left( {\partial _\zeta  \Theta } \right)^2  \\
\\
+ {\widetilde{{D}} ^{-1}}\left( {\partial _\vartheta \Theta } \right)^2, \\
 \end{array}
\end{eqnarray}
where ${W'}={\partial _r  S_0 } $ and $j={\partial _\phi  S_0 } $. For obtaining a nontrivial solution, the determinant of the matrix $\Lambda$ must equals to zero. Hence,
\begin{eqnarray}
\label{eq34}
{\rm{Im}} \tilde W_ \pm   =  \pm \int {\sqrt {\frac{{\tilde C\tilde D\tilde E\omega ^2  + \tilde X}}{{\tilde A\tilde B\tilde C\tilde D\tilde E}}} } dr =  \pm {{\pi r_H \omega } \mathord{\left/
 {\vphantom {{i\pi r_H \omega } 2}} \right.
 \kern-\nulldelimiterspace} 2},
\end{eqnarray}
where $ \tilde X=  - \tilde A\tilde D\tilde E\left( {\partial _\zeta  \Theta } \right)^2  - \tilde A\tilde C\tilde E\left( {\partial _\vartheta  \Theta } \right)^2  - \tilde A\tilde C\tilde Dj^2  - \tilde A\tilde C\tilde D\tilde Em^2$. The plus (minus) sign corresponds to the outgoing (incoming) solutions of vector particles. As a result, the tunneling rate is
\begin{eqnarray}
\label{eq35}
\begin{array}{l}
 \tilde \Gamma  = \frac{{\tilde \Gamma _{(emission)} }}{{\tilde \Gamma _{(absorpation)} }} =e^{ { - 4 {\mathop{\rm Im}\nolimits} W_ + } } =
e^{ { - 2\pi r_H \omega } }.\\
 \end{array}
\end{eqnarray}
With the formula $\Gamma  = \exp ({E \mathord{\left/ {\vphantom {E T}} \right. \kern-\nulldelimiterspace} T})$, where $E$ and  $T$ are  the energy of emitting particle and the temperature, we can calculate the Hawking temperature of 5-dimensional ST black hole
\begin{eqnarray}
\label{eq36}
\widetilde{T_H } =\frac{1}{{2\pi r_H }} = \frac{1}{{2\pi \sqrt M }}.
\end{eqnarray}
 The Hawking temperature of the 5-dimensional ST black hole is only related to the mass. People can get the same result when they investigate the scalar particles tunneling and Dirac particles tunneling from the 5-dimensional ST black hole. When assuming $D=4$, we find that the Hawking temperature of 4-dimensional ST black hole is $T_{H\left( {4D - ST} \right)}  = {1 \mathord{\left/ {\vphantom {1 {4\pi r_{H\left( {4D - ST} \right)} }}} \right.
 \kern-\nulldelimiterspace} {4\pi r_{H\left( {4D - ST} \right)} }}$ with ${r_{H\left( {4D - ST} \right)} } = 2M$, which is different from Eq. (\ref{eq36}).
This difference is caused by the property of ST black hole spacetime, it indicates that people may obtain different information from higher dimensional black hole.
\section{Discussion and conclusion}
In this paper, we studied the vector particles tunneling from 4-dimensional KdS black hole and 5-dimensional ST black hole. The tunneling rates and Hawking temperatures were gotten. For the 4-dimensional KdS black hole, we found that the tunneling rates and Hawking temperatures are not only depended on the outer event horizon, mass and angular momentum of the 4-dimensional KdS black hole but also the cosmological constant. Besides, Eq. (\ref{eq24}) and Eq. (\ref{eq25}) are consistent with that obtained by scalar particles and Dirac particles tunneling from the 4-dimensional KdS black hole. In the static limit, Eq. (\ref{eq25}) is reduced to the temperature of SC black hole. For 5-dimensional ST black hole, the $\widetilde{\Gamma}$ and $\widetilde{T_H}$ are related to the mass of the black hole. Moreover, since the $f(r)$ and the event horizon of ST black hole are depended on the dimension of spacetime, they lead the temperature of 5-dimensional ST black hole is different from the temperature of 4-dimensional ST black hole.

Moreover, by applying the WKB approximation, we derived the Hamilton-Jacobi equation from Proca equation.  \cite{ch19} and \cite{ch55} also derived the Hamilton-Jacobi equation from Klein-Gordon equation, Dirac equation and Rarita-Schwinger equation. Therefore, we think the Hamilton-Jacobi is a fundamental equation in the semiclassical theory, which can help people to investigate the semiclassical Hawking radiation behavior.

The Eq. (\ref{eq25}) and Eq. (\ref{eq36}) showed the Hawking radiation is the black-body radiation, it indicates that the black holes will emit away all their particles as the black-body radiation, that is, the black hole lose all its information. In order to solve this problem, the results need to be modified. In our further work, we will take into account the conservation of energy and the self-gravitational interaction.

\vspace*{3.0ex}
{\bf Acknowledgements}
\vspace*{1.0ex}

This work is supported by the Natural Science Foundation of China (Grant No. 11573022 and No. 11205125), the Sichuan Province Science Foundation for Youths (Grant No. 2014JQ0040), and the Innovative Research Team in China West Normal University (Grant No. 438061).

\end{document}